\documentclass[aps,prb,twocolumn,floatfix,groupedaddress]{revtex4}

\usepackage{graphicx}
\usepackage{amssymb}
\usepackage{amsmath}

\newcommand{\veps}{\varepsilon}
\newcommand{\eps} {\tilde{\varepsilon}}
\newcommand{\tR} {\tilde{R}}
\newcommand{\tr} {\tilde{r}}
\newcommand{\tf} {\tilde{f}}
\newcommand{\tmu} {\tilde{\mu}}

\renewcommand{\eqref}[1]{Eq. (\ref{#1})}
\newcommand{\figref}[1]{Fig. \ref{#1}}

\begin{document}
\title{Stability of Binary Mixtures in Electric Field Gradients}
\author{Sela Samin}
% \email{samins@bgu.ac.il}
\affiliation{Department of Chemical Engineering and The Ilse Katz Institute for 
Nanoscale Science and Technology, Ben-Gurion University of the Negev,
84105 Beer-Sheva, Israel.}

\author{Yoav Tsori}
% \email{tsori@bgu.ac.il}
\affiliation{Department of Chemical Engineering and The Ilse Katz Institute for
 Nanoscale Science and Technology, Ben-Gurion University of the Negev,
84105 Beer-Sheva, Israel.}

\date{19/11/2009}

% Last edited: 19/11/2009 by SS

\begin{abstract}

We consider the influence of electric field gradients on the phase behavior of nonpolar binary
mixtures. Small fields give rise to smooth composition profiles, whereas large enough fields 
lead to a phase-separation transition. The critical field for demixing as well as the 
equilibrium phase separation interface are given as a function of the various system parameters. 
We show how the phase diagram in the temperature-composition plane is affected by electric 
fields, assuming a linear or nonlinear constitutive relations for the dielectric constant. 
Finally, we discuss the unusual case where the interface appears far from any bounding surface.

\end{abstract}

\maketitle

\section{Introduction}

The effect of gravitational and magnetic fields on the phase diagram of liquid mixtures is
quite small in
general due to the weak coupling of the field with the mixture's composition. The influence of
electric fields was studied extensively, extensively in geometries where the field is
uniform. Theoretical work predicted that for a binary mixture the upper critical solution
temperature $T_c$ is shifted upward by a small amount, of the order of milikelvins
\cite{landau,onuki1}. Experiments in low molecular weight liquids predominantly showed an
opposite shift of the same magnitude \cite{debye,beaglehole,orzechowski,wirtz}. Two
exceptions are \textcite{reich} who measured the cloud point temperature
of a polymer mixture and \textcite{Szalai2008} who predicted a
downward shift in the critical pressure of dipolar fluid mixtures in
uniform electric fields.

In a uniform field the shift in $T_c$ is proportional to the square of electric field
$E^2$ and $d^2\veps/d\phi^2$ --
the second derivative of the dielectric constant with respect to the mixture composition
\cite{onuki:021506,boker_mm2009,tsori_mm2006,tsori_rmp2009,comment1,stepanow_mm2007,
stepanow_pre2009}. 
However, in realistic systems, such as colloidal suspensions and microfluidic devices, the
electric field
varies in space due to the complex geometry. In such systems, electric fields on the order of $10^7$V/m naturally occur due to the small length-scales involved. Recently, we have shown
that 
a homogeneous mixture confined by curved charged surfaces undergoes a phase-separation
transition and two distinct domains of high and low compositions appear
\cite{efdemix,efips_jcp1}. The transition occurs when the surface charge (or voltage) exceeds a
critical value. In this paper we examine in detail numerically and analytically the location of
the transition. We also show how this transition affects the temperature-composition phase
diagram. We find that near $T_c$ the spatial variation of the field leads to a nontrivial
modification of the stability lines. Our analysis shows that in nonuniform electric fields even
a linear constitutive relation can lead to a substantial change of the transition (binodal)
temperature.  Lastly, we demonstrate that the phase separation interface can appear far from
any of the surfaces bounding the mixture.

\section{Theory}

Consider an A/B binary mixture confined by charged conducting surfaces giving rise to an
electric field. The free energy of the mixture is 
\begin{equation}
\label{eq50}
F =\int {\left[ f_m (\phi,T)+f_{es}(\phi,\mathbf{E})\right] {\rm d} \mathbf{r}},
\end{equation}
where $f_m $ is the bistable mixing free energy density, given in terms of the A-component
composition $\phi$ ($0<\phi<1$) and the temperature $T$. 

$f_{es} $ is the electrostatic free energy density due to the electric field $\mathbf{E}$,
given by
\begin{equation}
\label{eq51}
f_{es} =-\frac{1}{2}\veps (\phi)(\nabla \psi)^2.
\end{equation}
$\psi$ is the electrostatic potential. The negative sign corresponds to cases where the
potential is prescribed on the bounding surfaces.  When the charge is given on the bounding
surfaces the sign should be replaced by a positive one
\cite{landau2,onuki_nato2004,onuki:021506}. $\veps$ is the mixture's
permittivity, and is a function of the composition $\phi$. Note that in order to isolate
the electric field effect 
we do not include any direct short- or long-range interactions
between the liquid and the confining surfaces. The equilibrium
composition profile $\phi(\mathbf{r})$ and electrostatic potential $\psi(\mathbf{r})$ are
given by the 
extremization of the free energy with respect to $\phi$ and $\psi$
\cite{onuki:3143,onuki_book}. 
The resulting Euler-Lagrange equations are
\begin{eqnarray}
\label{eq_gov1}
\frac{\delta F}{\delta \phi}&=&\frac{\delta f_m}{\delta \phi }-\frac{1}{2}\frac{d\veps(\phi)
}{d\phi }(\nabla \psi )^{2}-\mu=0,\\
\label{eq_gov2}
\frac{\delta F}{\delta \psi}&=&\nabla \cdot (\veps (\phi )\nabla\psi)=0.
\end{eqnarray}
\eqref{eq_gov2} is simply Gauss's law. Notice that the relation $\veps(\phi)$ couples these
nonlinear equations. In the canonical ensemble $\mu$ is a Lagrange multiplier adjusted to
satisfy mass conservation:
\begin{equation}
\label{eq_mass}
\int {\left[\phi(\mathbf{r})-\phi_0\right]  {\rm d} \mathbf{r}}=0 ,
\end{equation}
where $\phi_0$ is the average composition. In the case of a system in contact with a matter
reservoir (grand-canonical ensemble), the chemical potential is set by the reservoir, 
$\mu =\mu(\phi _0)$ where $\phi _0$ is the reservoir composition.

\begin{figure}[!t]
\includegraphics[width=3in]{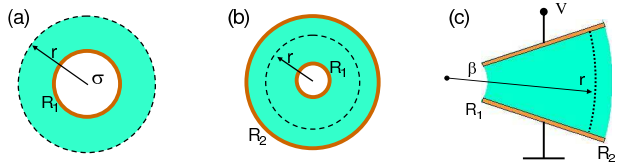}
\caption{The three model systems. (a) A single charged spherical colloid with surface charge
density $\sigma$ and radius $R_1$. (b) A charged wire with surface charge density $\sigma$ and
radius $R_1$, or two concentric cylinders with radii $R_1$ and
$R_2$. (c) A wedge comprised of two flat electrodes with an
opening angle $\beta$ and potential difference $V$. $R_1$ and $R_2$ are the minimal and
maximal values of the distance $r$ from the
imaginary meeting point of the electrodes.}
\label{fig_systems}
\end{figure}

In order to simplify the solution of Eqs. (\ref{eq_gov1})--(\ref{eq_mass}), we consider the 
three simple model systems shown schematically in \figref{fig_systems}. The first one is a 
charged isolated spherical colloid of radius $R_1$ and surface charge density $\sigma$, 
immersed in an infinite mixture bath. In this system, spherical symmetry dictates that  
$\phi=\phi(r)$ and $\mathbf{E}=\mathbf{E}(r)$ where r is the distance from the colloid's center. Since the colloid has a prescribed charge we 
can integrate Gauss's law and obtain an explicit expression for the electric field: 
$\mathbf{E}(r)=\sigma R_1^2/(\veps(\phi)r^2)\hat{\mathbf{r}}$. The second geometry is a charged wire of radius $R_1$ and surface charge 
density $\sigma$, coupled to a reservoir at $r \rightarrow \infty$.
Alternatively, we may consider a closed condenser made up of two concentric cylinders of radii 
$R_1$ and $R_2$. In both cases, we readily obtain the electric field 
$\mathbf{E}(r)=\sigma R_1/(\veps(\phi)r)\hat{\mathbf{r}}$, where $r$ is the distance from the 
inner cylinder's center. The last system is the wedge condenser, made up from two flat 
electrodes with an opening angle $\beta$ and a potential difference $V$ across them. Solution 
of the Laplace equation gives $\mathbf{E}(r)=(V/\beta r) \hat{\mathbf{\theta}}$, where $r$ is 
the distance from the imaginary meeting point of the electrodes and $\theta$ is the azimuthal 
angle. In this geometry $\phi=\phi(r)$ and therefore $\mathbf{E}\cdot \nabla\veps=0$. The 
explicit expressions for the electric field in all three systems outlined above decouple 
equations (\ref{eq_gov1}) and (\ref{eq_gov2}). 

We will show that \eqref{eq_gov1} leads, under certain conditions, to a phase-separation 
transition. This transition is independent of the exact form of $f_m$, and can be realized as 
long as $f_m$ is bistable and the dielectric constant $\veps$ depends on the composition 
$\phi$. In order to be specific, we will consider the mixing free energy derived from the 
Flory-Huggins lattice theory, with lattice site volume $v_0$. We consider the simple symmetric 
case where each component occupies $N$ successive lattice cells. Simple liquids have $N=1$, 
while polymers have $N>1$ monomers. The mixing free energy density is then given by 
$f_m=k_BT\tf_m/Nv_0$, where
\begin{eqnarray}\label{fm}
\tf_m=\phi\log(\phi)+
(1-\phi)\log(1-\phi)+N\chi\phi(1-\phi).
\end{eqnarray}
$k_B$ is the Boltzmann constant, and $\chi\sim1/T$ is the Flory interaction parameter 
\cite{doi}. We limit ourselves to the case where $\chi>0$, leading to an Upper Critical 
Solution Temperature type phase diagram in the $\phi-T$ plane. In the absence of electric 
field, the mixture is homogeneous above the binodal curve $\phi_t(T)$, and phase separates 
into two phases having the binodal compositions $\phi_t$ below it. Below the binodal curve, 
but above the spinodal, given by $\phi_{s}(T)=(1/2)[1\pm\sqrt{1-2/(N \chi)}]$, the mixture is 
metastable. The binodal and spinodal curves meet at the critical point 
$(\phi_c,(N\chi)_c)=(1/2,2)$. The transition (binodal) temperature $T_t$ for a given 
composition is given by 
$T_t(\phi)=(N\chi)_cT_c\left[\log(\phi/(1-\phi))/(2\phi-1)\right]^{-1}$ \cite{doi}.

Using the expressions given above for the electric field, we write the generalized composition 
equation valid for cylindrical and spherical geometries:

\begin{eqnarray}\label{gov_eqn1}
\tf'_m(\phi)-
N\chi M_{{\rm sc}}\frac{d\eps/d\phi}{\eps^2(\phi)}\tr^{-n}-\tmu=0.
\end{eqnarray}
Here, $\tr\equiv r/R_1$ is the scaled distance and $\eps=\veps/\veps_0$, with $\veps_0$ the
vacuum permittivity. Where
\begin{eqnarray}
M_{{\rm sc}}\equiv \frac{\sigma^2Nv_0}{4k_BT_c\veps_0}
\end{eqnarray}
is the dimensionless field, and $n$ is an exponent characterizing the decay of $E^2$: $n=2$ 
for concentric cylinders, and $n=4$ for spherical colloid. For the wedge geometry we find:
\begin{eqnarray}\label{gov_eqn2}
\tf'_m(\phi)-
N\chi M_{{\rm w}}\frac{d\eps}{d\phi} \tr^{-n}-\tmu=0,
\end{eqnarray}
where
\begin{eqnarray}
M_{{\rm w}}\equiv \frac{V^2Nv_0\veps_0}{4\beta^2k_BT_cR_1^2},
\end{eqnarray}
and $n=2$.
$M_{{\rm sc}}$ and $M_{{\rm w}}$ are dimensionless quantities measuring the magnitude of the 
maximal electrostatic energy stored in a molecular volume compared to the thermal energy. The 
second term in Eqs. (\ref{gov_eqn1}) and (\ref{gov_eqn2}) is the variation of the electrostatic 
free energy with respect to $\phi$, and is only present when the mixture components have 
different permittivities.

The constitutive relation $\eps(\phi)$ is a smooth function of $\phi$. Experiments show
that the curve can be slightly convex or concave, and is dominantly linear
\cite{debye,beaglehole,Akhadov}. They are mostly in agreement with 
Clausius-Mossotti and Onsager-based theories for the dielectric constant
\cite{Sen1992}. Thus, for a mixture of liquids A and B with
dielectric constants $\eps_a$
and  $\eps_b$, respectively, the experiments yield a polynomial relation in the form
\begin{equation}\label{const_relation}
\eps(\phi)\simeq \eps_b+\eps'\phi+\eps''\phi^2+.~.~.~.
\end{equation}
We start by focusing on a linear relation, namely $\eps''=0$; in this case
$\eps'=\Delta\eps\equiv \eps_a-\eps_b$. Even in such a simple case it turns out that a
phase separation transition occurs, in contrast to the Landau mechanism which relies on a
nonvanishing $\eps''$. After investigating linear relations we examine how our results
change when $\eps(\phi)$ has a positive or negative curvature, by allowing for $\eps''\neq
0$. In this case $\eps'$ is different from $\Delta \eps$. 
Higher order terms in the expansion Eq. (\ref{const_relation}) are not expected to
change the results qualitatively, since they do not affect much the curvature of
$\eps(\phi)$\cite{debye,beaglehole,Akhadov}.
 
 \begin{figure}[!tb]
%fig : /users/sela/work/efips/newQ/efips_#2_figs
%plot : plot_graph_sol_wedge.m
\centerline{\includegraphics[width=3in,clip]{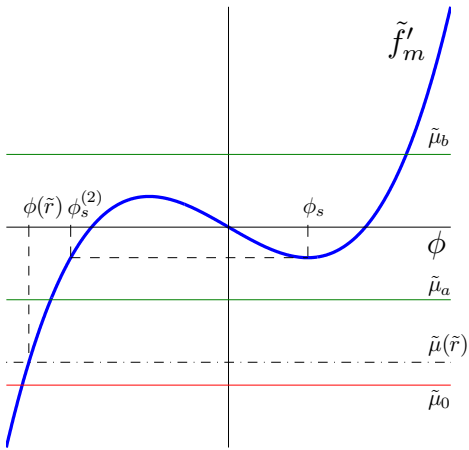}}
\caption[Graphical solution of the composition equation for an open wedge]{Graphical solution 
of \eqref{wo_eqn} for an open wedge at $T<T_c$ and a symmetric mixture. Solid curve is 
$\tf'_m(\phi)$. Its roots are the transition (binodal) compositions. The intersection between 
$\tf'_m(\phi)$ and the horizontal dash-dotted line $\tmu(\tr)$ gives the composition 
$\phi(\tr)$. If $\tmu(\tR_1)$ is at $\tmu_a$, the profile $\phi(\tr)$ varies smoothly, but if 
$\tmu(\tR_1)=\tmu_b$, $\phi(\tr)$ has a discontinuity.}
\label{fig_graph_sol}
\end{figure}

\section{Results and Discussion}
Before we present the numerical solutions of Eqs. (\ref{gov_eqn1}) and (\ref{gov_eqn2}), it is 
illustrative to consider a graphical solution for a wedge condenser. Recall that in the 
absence of field it is assumed that $T$ is above the binodal temperature. We rewrite 
\eqref{gov_eqn2} as:
\begin{equation}\label{wo_eqn}
\begin{split}
\tf'_m(\phi)&=\tmu(\tr)\\
\tmu(\tr)&\equiv\frac{N\chi\Delta\eps M_{{\rm w}}}{\tr^2}+\tmu_0
\end{split}
\end{equation}
where $\tmu_0$ is the dimensionless reservoir chemical potential. At a given temperature, the intersection of
$\tf'_m(\phi)$ and the horizontal line $\tmu(\tr)$ gives the local  composition $\phi(\tr)$
(\figref{fig_graph_sol}). When $\tr\rightarrow\infty$, $\tmu(\tr)\rightarrow\tmu_0$ and the
composition is $\phi=\phi_0$, corresponding to a homogeneous phase. For simplicity we consider
$\phi_0<\phi_c$. As $\tr$ decreases, $\tmu(\tr)$ (and hence $\phi(\tr)$) increase until they
attain their maximal value at $\tR_1$. Above $T_c$, the free energy is always convex, $\tf'_m$
is a monotonic function of $\phi$, and the composition profile $\phi(\tr)$ is hence continuous.
However, below $T_c$, $\tf_m$ is bistable and $\tf'_m$ is sigmoidal. In this case there are two
possible scenarios shown in \figref{fig_graph_sol}. If, for example, $\tmu(\tR_1)=\tmu_a$, the
composition profile varies smoothly. If, on the other hand, $\tmu(\tR_1)=\tmu_b$, there is a
discontinuity in the profile since there is a radius $\tR>\tR_1$ where the value of $\phi$ can
``jump'' from high to low values. Below $T_c$, there is a range of radii, or compositions,
where the discontinuity in $\phi(\tr)$ can occur. The equilibrium profile $\phi(\tr;\tR)$ is
the one that minimizes the total free energy integral $F=\int f[\phi(\tr;\tR)]{\rm d}r$. 
These conclusions also hold for \eqref{gov_eqn1}. 

In \figref{fig_graph_sol} we show the composition $\phi_s^{(2)}$ defined by the relation 
$\tf'_m(\phi_s^{(2)})=\tf'_m(\phi_{s})$. Clearly, this is the minimal composition for which 
exist more than one solution to \eqref{gov_eqn2}. The role of this special composition will 
be discussed later.

\begin{figure}[!tb]
%fig : /users/sela/work/efips/newQ/efips_#2_figs
%plot : plot_efips_prof_types.m
\centerline{\includegraphics[width=3.5in,clip]{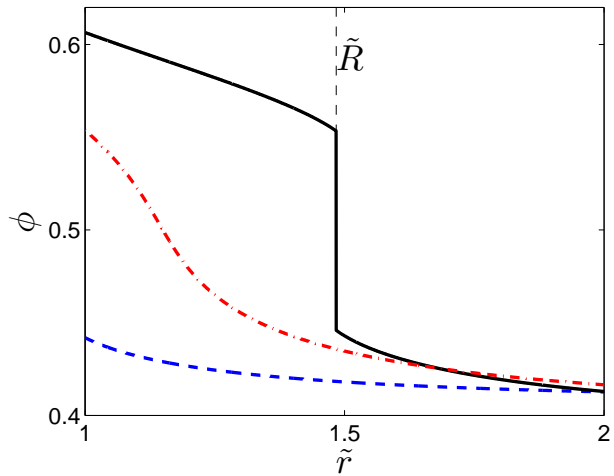}}
\caption{The three types of equilibrium profiles $\phi(\tr)$ for a system of two concentric 
cylinders. Dash-dot line: $T=1.005T_c$ and $M_{{\rm sc}}=0.04$: above $T_c$ the profile is 
smoothly varying. Dashed line: $T=0.995T_c$, $M_{{\rm sc}}=0.01$ smaller than the critical 
value for demixing. Solid line: same $T$, but $M_{{\rm sc}}=0.04$ is large enough to induce 
phase separation marked by an interface at $\tr=\tR$. We took an average composition 
$\phi_0=0.41$. Here and in other figures, $\tR_1=1$, $\tR_2=5$, $\eps_a=5$, and $\eps_b=3$.}
\label{fig_prof_type}
\end{figure}

The three typical composition profiles are shown in \figref{fig_prof_type}. Above $T_c$ 
(dash-dot line), $\phi(\tr)$ varies smoothly due to the dielectrophoretic force, whereby the 
high-$\eps$ liquid is drawn into the strong electric field region. Below $T_c$ (dashed line), 
at a temperature where the field-free mixture is homogeneous, if $M$ is small the profile 
$\phi(\tr)$ is again smoothly decaying, exhibiting the same dielectrophoretic behavior. 
However, if at the same temperature, $M$ is increased, either by adding charge to the surface 
or by increasing the curvature (smaller $R_1$), we arrive at a critical value, denoted $M^*$. 
Above it, a phase-separation transition occurs. This is shown in the solid line of 
\figref{fig_prof_type}, where the mixture consists of two coexisting domains separated by an interface at $\tr=\tR$.

The typical value of charge/voltage required for demixing can be estimated from the value of 
$M$ being in the range $M\sim 0.001-0.1$ \cite{efips_jcp1}. Consider a colloid of radius 
$R_1\sim1\mu$m placed in a mixture having a molecular volume $Nv_0\simeq5\times 10^{-27}$m$^3$ 
and $T_c\simeq 300$K. Then, the typical demixing charge is of the order of $10^2-10^4e$ 
charges (surface voltage is $1-100V$). It scales linearly with $1/N$: in a polymer mixture the 
confining surfaces require $N$ times smaller charge compared to molecular liquids in order to 
induce phase separation. 

\subsection{The phase-separation interface}

\begin{figure}[!t]
\begin{center}
%fig : /users/sela/work/efips/newQ/plot_figR2.m
\includegraphics[width=3in,clip]{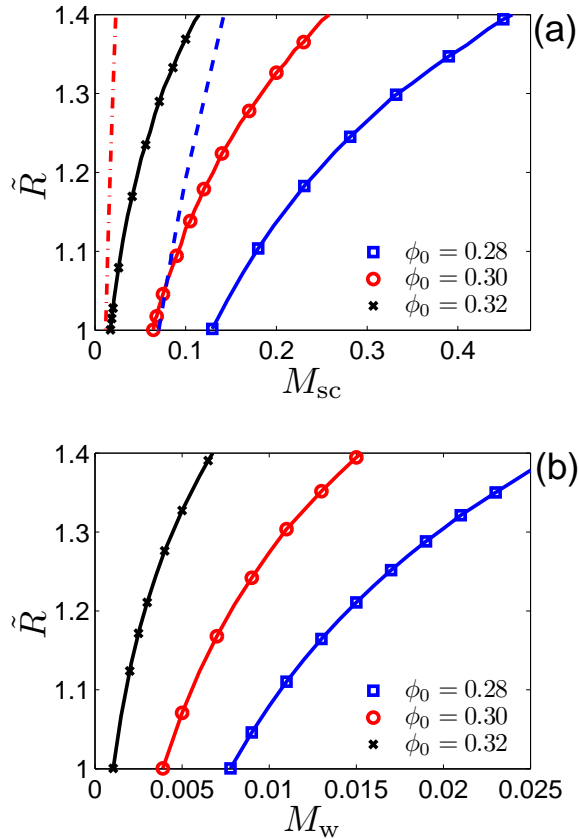}
\end{center}
\caption{Location of the demixing interface $\tR$ as a function of $M$, (a) for three average 
compositions $\phi_0$ in a closed cylindrical system (lines with symbols) and for 
$\phi_0=0.28$ (dashed line) and $\phi_0=0.3$ (dash-dot line) in an open system. Curves do not 
coincide at $\tR=1$. (b) The same, in the closed "wedge" geometry. We took $T=0.96T_c$.}
\label{fig_RvM}
\end{figure}

At the critical value of $M$, a sharp interface first appears separating coexisting regions of 
high- and low-$\phi$ value. If the average composition $\phi_0$ is smaller than $\phi_c$, the 
interface appears at $\tr=\tR_1$. When we further increase $M$ and supply more electrostatic 
energy to the system, dielectrophoresis leads to an increase in the size of the high 
composition domain. Thus, the location of the separation interface $\tR$ increases. 
\figref{fig_RvM} shows how $\tR$ varies with $M$ at a constant temperature in the concentric 
cylinders and wedge systems. Notice that as $\phi_0$ approaches the binodal composition 
($\phi_t\simeq0.33$), $\tR$ is larger at the same $M$. It also grows more rapidly with 
increasing $\phi_0$. Indeed, when the binodal is approached, the mixing free energy barrier is 
smaller. Secondly, $\tR$ grows faster in an open system than in a closed one. This is because 
in a closed system the energy penalty in $\tf_m$ grows faster than the energy gain in 
$\tf_{es}$. Material conservation gives the maximum value of $\tR$, $\tR_{\infty}$, given by 
\begin{equation}
\tR^2_{\infty}=\phi_0(\tR_2^2-\tR_1^2)+\tR_1^2.
\label{eq:rinf}
\end{equation}
This $M\rightarrow\infty$ limit is physically unattainable and is preempted by dielectric 
breakdown.

\begin{figure}[!t]
\begin{center}
%fig : /users/sela/work/efips/newQ/plot_figR3.m
\includegraphics[width=3in,clip]{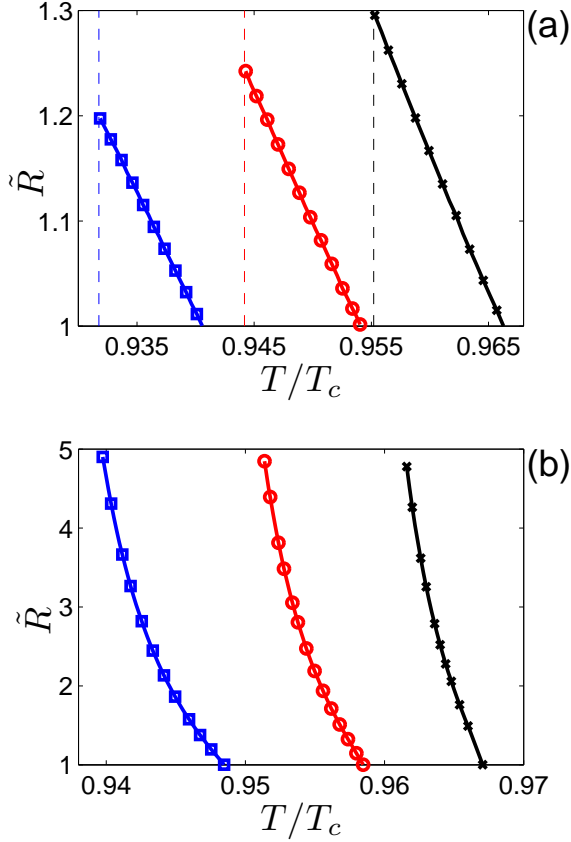}
\end{center}
\caption{Location of the demixing interface $\tR$ as a function of $T/T_c$, (a) for three 
average compositions in a closed cylindrical system: $\phi_0=0.28$ (squares), $0.3$
(circles) and $0.32$ (crosses) with $M_{{\rm sc}}=0.04$. When $\tR=\tR_1$, the temperature 
corresponds to $M_{{\rm sc}}=M_{{\rm sc}}^*=0.04$. $\tR$ grows as $T/T_c$ is reduced until
it 
attains it maximal value at the binodal temperature (dashed line for each value  of
$\phi_0$) 
(b) Same, in an open cylindrical system with $M_{{\rm sc}}=0.004$. Here, $\tR \rightarrow 
\infty $ at the binodal.
}
\label{fig_RvT}
\end{figure}

Note that the typical values of $M$ in the cylindrical and spherical cases are an order of 
magnitude larger than in the wedge condenser, see \figref{fig_RvM} (a) and (b). Indeed, in the 
spherical and cylindrical symmetries,
$\mathbf{E}$ is parallel to $\nabla \phi$: the dielectrophoretic
force, (proportional $\Delta \veps$), has to be large enough to overcome the energy penalty
associated with dielectric interfaces parallel to $\mathbf{E}$ (proportional to $(\Delta
\veps)^2$) \cite{tsori_rmp2009}. On the other hand, in the wedge condenser $\mathbf{E}$ is
perpendicular to $\nabla \phi$, and the required dielectrophoretic force for demixing is
correspondingly smaller, leading to smaller values of $M_{{\rm w}}$. This could be seen by
comparing the electrostatic terms in Eqs. (\ref{eq_gov1}) and (\ref{eq_gov2}), differing by a
factor of $\eps(\phi)^{-2}\sim0.1$, which $M_{{\rm sc}}$ has to compensate for in order for the
values of $\tf_{es}$ to be equal.

Alternatively, an increase in $T$ at constant electric field decreases $\chi M\sim M/T$ and 
decreases $\tR$.
\figref{fig_RvT} shows how an increase in $T$ shrinks the high-$\phi$ domain and decreases 
$\tR$ in a closed and open cylindrical system. In \figref{fig_RvT}, when $\tR=1$ the 
temperature is that for which $M=M^*$. The effect of temperature is much more pronounced in an 
open system: $\tR$ tends to infinity when approaching the binodal temperature (not shown). On 
the other hand, $\tR$ is finite when approaching the binodal in a closed system. Its maximal 
value is larger when $|\phi_0-\phi_c|$ is smaller, because then the mixing free energy 
difference between low- and high-$\phi$ values is reduced. In \figref{fig_RvT} (a), $\tR$ 
appears to be linear simply because $T$ changes over a small interval.

\begin{figure}[!t]
\begin{center}
%data source : /users/sela/work/efips/newQ/newopen/getimspintrans.m
%fig : /users/sela/work/efips/newQ/efips_#2_figs
%data : efips_imspintrans_wedge.mat
%plot : plot_efips_imspintrans_wedge.m
%file : efips_imtrans_inset.fig
\includegraphics[width=3.5in,clip]{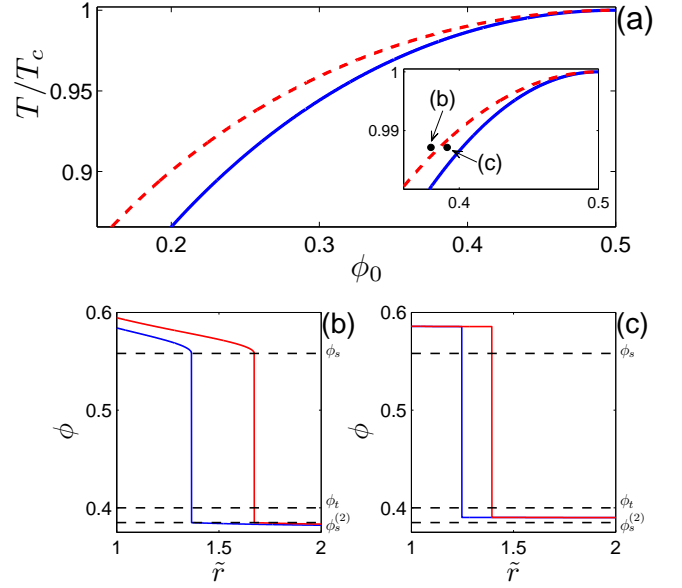}
\end{center}
\caption[The ``differentiating curve'' for an open wedge]{(a) The ``differentiating'' curve --
$\phi^\dagger$ (dashed line) for an open wedge. Above it, when phase separation occurs, the
composition at the interface jumps from $\phi_s^{(2)}$ to $\phi_s$. Below $\phi^\dagger$ the
front composition jumps from $\phi_s^{(2)}<\phi_1<\phi_t$ to $\phi_2>\phi_s$. Examples of this
behavior are shown in (b) and (c), where composition profiles for points above and below
$\phi^\dagger$ with two values of $M$ are given, showing this behavior is independent of $M$.
The inset in (a) is a blowup showing the location of (b) and (c) in the $\phi_0-T$ plane. In
(b) $\phi_0=0.38$ with $M_{{\rm w}}=4\times 10^{-4}$ and $M_{{\rm w}}=6\times 10^{-4}$, and in
(c) $\phi_0=0.39$ with $M_{{\rm w}}=1\times 10^{-5}$ and $M_{{\rm w}}=1.5\times 10^{-5}$. In
(b) and (c) we took $T=0.987T_c$.
}
\label{fig_dummy}
\end{figure}

One can estimate the value of $\phi$ at the demixing interface in an open system. At the
interface there is a ``jump'' in $\phi(\tr)$ from $\phi_1$ to $\phi_2>\phi_1$. Let us denote by
$\phi(\tR)$ the upper interface composition -- $\phi(\tR)=\phi_2$. The conditions for a
``jump'' in the wedge geometry are:
\begin{eqnarray}
\label{condw1}
\tf_m'(\phi_1)- N\chi \frac{\Delta \eps M_{{\rm w}}}{\tR^2}-\tmu_0=0,\\
\label{condw2}
\tf_m'(\phi_2)- N\chi \frac{\Delta \eps M_{{\rm w}}}{\tR^2}-\tmu_0=0,\\
\label{condw3}
\tf_m(\phi_1)- N\chi \frac{\eps(\phi_1)M_{{\rm w}}}{\tR^2}\geq \tf_m(\phi_2)- N\chi 
\frac{\eps(\phi_2)M_{{\rm w}}}{\tR^2}.
\end{eqnarray}
The first two equations define the local solutions of \eqref{wo_eqn}, and
the third one is the condition that a high composition is favorable: 
$\tf(\phi_2)<\tf(\phi_1)$. The true value of $\tR$ is the one that gives the global minimum of 
the free energy. Putting \eqref{condw1} in \eqref{condw3} and using
$\tmu_0=\tf_m'(\phi_0)$, we 
get
\begin{eqnarray}
\label{condwj}
\tf_m'(\phi_1)-\tf_m'(\phi_0) \geq \frac{\tf_m(\phi_2)-\tf_m(\phi_1)}{\phi_2-\phi_1}.
\end{eqnarray}
The right-hand side of \eqref{condwj} is maximal when the transition occurs from 
$\phi_1=\phi_s^{(2)}$ to $\phi_2=\phi_{s}$ ($\phi_s^{(2)}<\phi_{s}$). We therefore denote 
$\Delta f_{{\rm w, max}}$ by
\begin{eqnarray}
\Delta f_{{\rm w, max}}=\frac{\tf_m(\phi_{s})-\tf_m(\phi_s^{(2)})}{\phi_{s}-\phi_s^{(2)}}. 
\end{eqnarray}
If the inequality
\begin{eqnarray}
\label{condwdspin}
\tf_m'(\phi_1)-\tf_m'(\phi_0) \geq \Delta f_{{\rm w, max}}
\end{eqnarray}
holds, the transition must be at $\phi_1=\phi_s^{(2)}$, since for larger values of  $\phi_1$ 
the right-hand side of \eqref{condwj} is smaller while the left-hand side is larger, so a 
higher composition is surely favored. The equality sign in \eqref{condwdspin} corresponds to 
the maximal average composition $\phi_0$ for which this equation holds. 

 \begin{figure}[!t]
\begin{center}
%fig : /users/sela/work/efips/newQ/efips_#2_figs
%data : efips_phi_trans_closed.mat 
%plot : plot_efips_phi_trans_closed.m
\includegraphics[width=3in,clip]{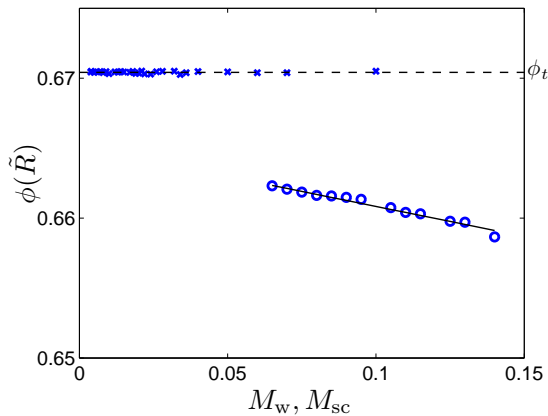}
\end{center}
\caption{Composition at the demixing interface $\tR$ as a function of $M$ for a closed wedge 
(crosses) and concentric cylinders (circles). The dashed line is the binodal composition. Here 
$\phi_0=0.3$ and $T=0.96T_c$.}

\label{fig_jumpc}
\end{figure}

The locus of such compositions is the ``differentiating curve'' -- $\phi^\dagger$, shown in the
dashed curve of \figref{fig_dummy} (a). When the zero-field point in the phase diagram
$(\phi_0,T)$ is above the $\phi^\dagger(T)$ curve, \eqref{condwdspin} holds, and the
composition at the interface jumps from $\phi_s^{(2)}$ to $\phi_s$ \cite{comment2}. 
When $(\phi_0,T)$ is below
$\phi^\dagger(T)$ the upper interface composition $\phi(\tR)$ is between $\phi_s$ and $\phi_t$.
\eqref{condwj} shows this result is independent of $M$. An example of this situation is given
in \figref{fig_dummy} (b), where the compositions $\phi_1$ and $\phi_2$ are the same for two
values of $M$. \figref{fig_dummy} (c) shows the interface compositions are independent of $M$
also when $\phi_0$ is larger than $\phi^\dagger$. 

When the discontinuity in the composition profile occurs at $\phi_2=\phi_s$, we can invert 
\eqref{gov_eqn2} to get $\tR\propto M^{1/2}$ for the open wedge system. In particular, above 
the differentiating curve we get in the Flory-Huggins model
\begin{eqnarray}\label{profile}
\label{rtexp}
\tR=\left[
\frac{N\chi \Delta \eps M_{{\rm w}}}{\tf_m'(\phi_s)-\tmu_0}
\right]^{1/2}.
\end{eqnarray} 
Thus, $\tR$ varies linearly with the wedge potential $V$.

In the other geometries $\tf'_{es}$ depends on $\phi$, and this influences the value of pairs 
$\phi_1$ and $\phi_2$. However, a very good approximation, valid when $M_{{\rm sc}}$ is not 
too large and $T$ is not too close to $T_c$, is that the transition remains from 
$\phi_1=\phi_s^{(2)}$ to $\phi_2=\phi_{s}$. One can then repeat a similar derivation and obtain
\begin{eqnarray}
\tf_m'(\phi_1)-\tf_m'(\phi_0) \geq \Delta f_{{\rm sc, max}},\\
\Delta f_{{\rm sc, 
max}}=\frac{\tf_m(\phi_{s})-\tf_m(\phi_s^{(2)})}{\phi_{s}-\phi_s^{(2)}}\frac{\eps(\phi_{s})}
{\eps(\phi_{s}^{(2)})}. 
\end{eqnarray}
Using these equations one can determine the differentiating curve for cylindrical and 
spherical geometries.

The situation is different in a closed system, as \figref{fig_jumpc} shows. In the wedge, the 
demixing interface occurs at the binodal composition, $\phi_t$, irrespective of $M$. In the 
cylindrical and spherical systems, $\phi(\tR)$ is lower than but close to $\phi_t$, and 
decreases when $M$ grows. The qualitative explanation is as follows. In the wedge geometry 
$\tf'_{es}$ only adds a constant to $\tf'$, and the binodal compositions remain the only pair 
of solutions of \eqref{gov_eqn2} that have the same mixing free energy $\tf_m$. Hence, the 
mixing free energy penalty is minimized when the transition is at the binodal compositions 
\cite{landau3}. This also explains why in the wedge geometry $\phi(\tR)$ is independent of 
$M$. In the cylindrical and spherical geometries on the other hand, $\tf'_{es}$ affects 
$\phi(\tR)$, resulting in a value of $\phi(\tR)$ smaller than $\phi_t$. This reflects the fact 
that dielectrophoresis favors high values of $\phi$. Since $\tf_{es} \propto M$, larger values 
of $M$ lead to lower values of $\phi(\tR)$.

\begin{figure}[!t]
%data source : /users/sela/work/efips/newQ/newopen/plot_pdo.m, 
%/users/sela/work/concyl/newQ/plot_pd.m
%fig : /users/sela/work/efips/newQ/efips_#2_figs
%data : efips_pdo_sph.mat, efips_pdc_cyl.mat
%plot : plot_efips_pd_lindeps.m
\begin{center}
\includegraphics[width=3.5in,clip]{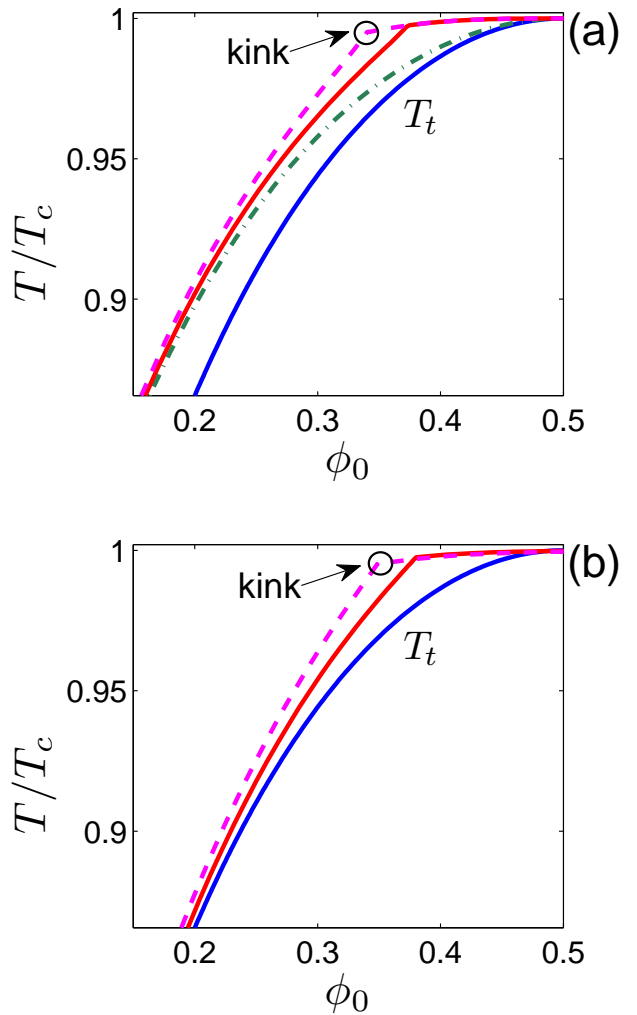}
\end{center}
\caption[Stability diagram showing $\phi^*(T)$ for a spherical colloid and concentric 
cylinders]{(a) Stability diagram of a spherical colloid with $M_{{\rm sc}}=0.04$ (thick solid 
line) and $M_{{\rm sc}}=0.08$ (dashed line). Dash-dot curve is $\phi^\dagger$ [see 
\figref{fig_dummy}(a)] and thin solid line is the binodal $T_t(\phi_0)$ . (b) Stability 
diagram showing $\phi^*(T)$ for concentric cylinders with $M_{{\rm sc}}=0.04$ (thick solid 
line) and $M_{{\rm sc}}=0.08$ (dashed line).
}
\label{fig_sd1}
\end{figure}

\subsection{Stability diagrams}

One can also set $M$ constant and for a given electric field draw the a stability curve 
$\phi^*(T)$ in the $\phi_0-T$ plane, see \figref{fig_sd1}. $\phi^*(T)$ is defined such that 
below it phase separation occurs, while above it composition profiles are smooth.  
\figref{fig_sd1} (a) and (b) show the stability diagram of a spherical colloid and concentric 
cylinders, respectively. 
In both diagrams, an increase in $M$ increases the unstable region. For the same value of $M$, 
in a closed system the phase separation region is smaller than in an open one, because the 
mixing energy penalty makes it more difficult to induce phase separation. The range of values 
of $\phi_0$ that are unstable in nonuniform electric fields grows when $T$ increases (but 
still $T<T_c$). 
For low values of $T$, there is a significant difference between open and closed systems: in 
open systems, if $M$ is large enough the stability curve tends to $\phi^\dagger$ [see 
\figref{fig_dummy}(a)], whereas for closed systems the stability curve tends to $\phi_t$ where 
demixing is spontaneous.

In both parts of \figref{fig_sd1}, there is a kink in all the curves $\phi^*(T)$ at a 
temperature we denote $T_{k,1}$. In the Flory-Huggins model and for spherical colloids and 
concentric cylinders, the second derivative of the free energy is
\begin{eqnarray}\label{dfes2_eqn}
\frac{\partial^2 \tf}{\partial\phi^2}=\frac{\partial^2 \tf_m}{\partial\phi^2}+2N\chi M_{
{\rm sc}}\frac{(d\eps/d\phi)^2}{\eps^3(\phi)}\tr^{-n}.
\end{eqnarray}
The second term in this equation is the positive electrostatic contribution $f''_{es}$. It is 
clear that when an electric field is present, even at $T<T_c$, the electrostatic contribution 
can lead to a positive value of $\tf''(\phi,\tr)$ and phase separation cannot occur.

%location : /users/sela/work/efips/newQ/efips_#2_figs/tr_lt_inset2.fig
\begin{figure}[!tb]
\begin{center}
\includegraphics[width=3in,clip]{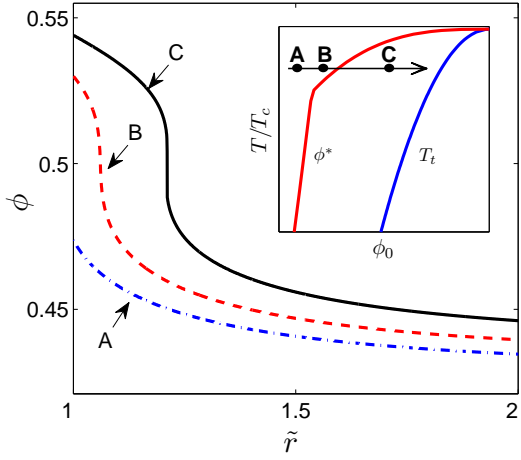}
\end{center}
\caption{Composition profiles above the kink temperature for an isolated spherical colloid. 
The profiles change from smooth (dash-dot and dashed lines) to discontinuous (solid line) with 
a discontinuity at finite value of $\tR$: $\tR_1\le\tR\le\tR_2$, when $\phi_0$ increases at 
constant $T$ and $M$ (see inset).
}
\label{fig_kink}
\end{figure}

\begin{figure}[!tb]
\begin{center}
\includegraphics[width=3in,clip]{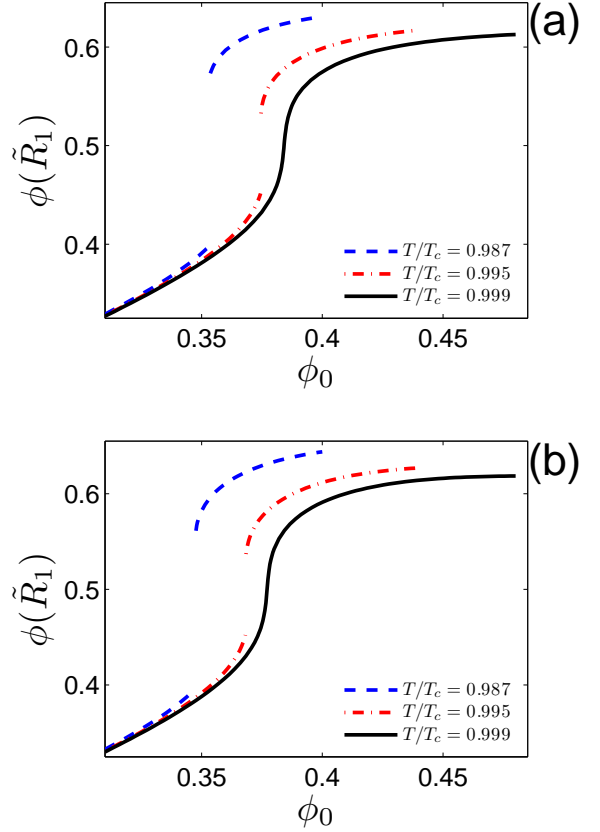}
\end{center}
\caption{Surface composition $\phi(\tR_1)$ when approaching the binodal, (a) for three 
temperatures in a closed cylindrical system. (b) Same, for an isolated charged cylinder. We 
took $M_{{\rm sc}}=0.04$.
}
\label{fig_phi1}
\end{figure}

When $T<T_{k,1}$, $\tf''$ is negative for all values of $\tr$, and the phase separation 
interface appears first at $\tR=\tR_1$. However, when $T>T_{k,1}$, for a given value of 
$\phi_0$, a special radius $\tR_c(T,\phi_0)$ exists. This is the largest value of $\tr$ for 
which $\tf''$ in \eqref{dfes2_eqn} can be negative. In this case, the demixing interface 
appears first at $\tR=\tR_c$.
An example of this behavior is shown in \figref{fig_kink}: at constant $T_c>T>T_{k,1}$, at 
points A and B ($\phi_0<\phi^*$) $\phi(\tr)$ is smooth, similar to $\phi(r)$ above $T_c$. 
However, at point C ($\phi_0>\phi^*$) $\phi(r)$ has a discontinuity at $\tR>\tR_c$.
As the critical point is approached, $\tR_c \rightarrow \infty$ and $\phi^*$ approaches the 
critical composition. 

\begin{figure}[!tb]
\begin{center}
%data source : /users/sela/work/efips/newQ/newopen/epsfun/plot_pdo_all.m, 
%/users/sela/work/concyl/newQ/epsfun/collect_phic.m
%fig : /users/sela/work/efips/newQ/efips_#2_figs
%data : efips_pdo_wedge_quadeps.mat, efips_pdc_wedge_quadeps.mat
%plot : plot_efips_pd_quadeps_full.m
\includegraphics[width=3.5in,clip]{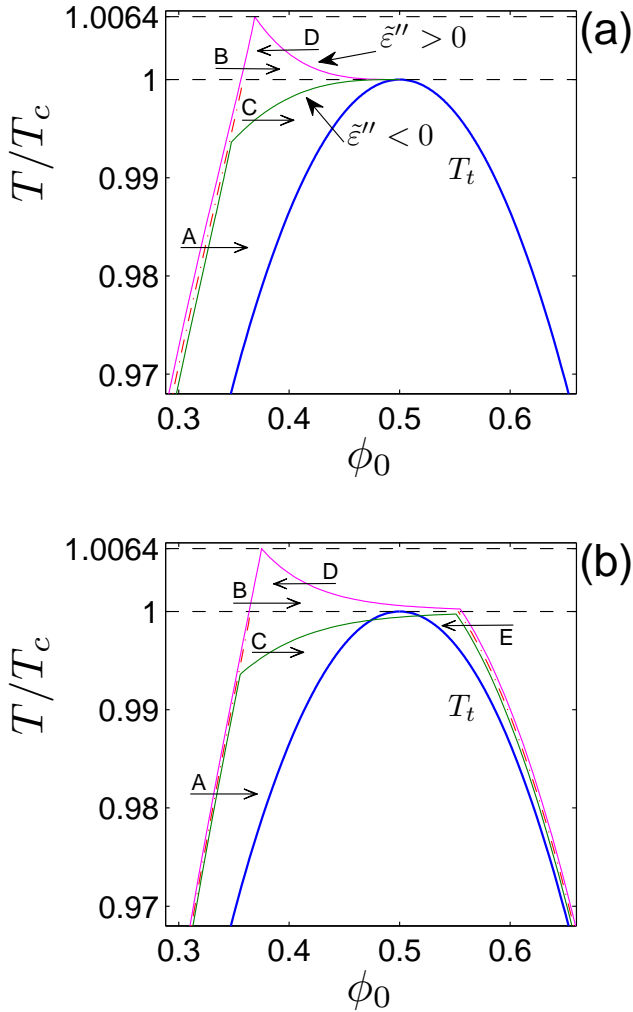}
\end{center}
\caption{(a) Stability diagram for an open wedge with $M_{{\rm w}}=4\times10^{-3}$ and three 
constitutive relations Eq. (\ref{const_relation}) with $\eps_a=5$ 
and $\eps_b=3$. Thick solid line is the transition (binodal) curve $T_t(\phi)$. When $\eps''=0$ 
(dash-dot line) there is no phase separation above $T_c$. For $\eps''=1.6$ an 
unstable region exists above $T_c$. In contrast, when $\eps''=-1.6$ is negative, 
the unstable region is suppressed. $\phi(\tr)$ changes from smooth to discontinuous at $\tR_1$ 
along lines A and B, and at $\tR_c>\tR_1$ along C and D. (b) The same for a closed wedge 
system. Note that the stability curve extends to compositions larger than $\phi_c$.  
$\phi(\tr)$ changes from smooth to discontinuous at $\tR_1$ along A and B, at 
$\tR_1<\tR_c<\tR_2$ along C and D, and at $\tR_2$ along E. Here $T_{k,1}/T_c=1.0064$.
}
\label{fig_sd2}
\end{figure}

The kink temperature $T_{k,1}$ is given by setting $\tR_c=\tR_1$ and can be obtained from 
solution of:
\begin{eqnarray}\label{eqn_pTk}
\frac{\partial^2  \tf}{\partial \phi^2}(\tR_1)=0,\  \frac{\partial^3 \tf}{\partial 
\phi^3}(\tR_1)=0.
\end{eqnarray} 
We stress that this result is independent of the exact form of mixing free energy. 
Notice that for a wedge, the electric field has no effect on the convexity of the free energy, 
and one finds $T_{k,1}=T_c$ (see \figref{fig_sd2}).

The surface composition $\phi(\tR_1)$ when approaching the binodal at constant $T$ and $M$ is 
given in \figref{fig_phi1}. When $T<T_{k,1}$ the surface composition has a discontinuity at 
$\phi_0=\phi^*$ (dashed and dash-dot lines) and the value of $\phi(\tR_1)$ becomes larger than 
$\phi_c$.  When $T>T_{k,1}$ the surface composition varies smoothly (solid lines). The 
discontinuity in $\phi(\tR_1)$ occurs at lower values of $\phi_0$ in open systems compared to 
closed ones; open systems are less stable than closed systems. When $T$ increases at a given 
value of $\phi_0$, the surface composition decreases since mixing is favored at high 
temperatures.

\subsection{Quadratic constitutive relation}

We now examine how the stability diagram changes if the dielectric constant has a quadratic 
dependence on composition: $\eps''\neq 0$ in
Eq. (\ref{const_relation}).  For simplicity,
we 
treat the wedge system where a linear constitutive relation means that $T_{k,1}=T_c$. In the 
Flory-Huggins model, the conditions in \eqref{eqn_pTk} give 
\begin{eqnarray}\label{eqn_tk1}
\frac{T_{k,1}}{T_c}=1+M_{{\rm w}}\eps'',
\end{eqnarray} 
where we used $\tR_1=1$. 

Note that $\eps''$ can be positive or negative. When $\eps''<0$, $\tf''_{es}$ is positive 
and we return to the same behavior we saw for spherical and cylindrical systems with linear 
relation $\eps(\phi)$. On the other hand, if $\eps''>0$ then $f''_{es}$ is negative and phase 
separation is possible above $T_c$. The stability diagram of a wedge with a quadratic 
constitutive relation is shown in \figref{fig_sd2}. In this figure, arrows labeled A--E 
indicate the variation of $\phi_0$ at constant $T$ in different areas of the stability 
diagrams. For each arrow the location for which the interface first appears is given in the 
caption of \figref{fig_sd2}. For the data in \figref{fig_sd2}, the kink temperature is given 
by $T_k/T_c=1\pm0.0064$, depending on the sign of $\eps''$. Taking $T_c\approx300$K, we have 
$\Delta T(\phi^*(T_k))=T_k-T_c\approx\pm2$K. This change in $T_c$ is two orders of magnitude 
larger than the corresponding change in uniform electric fields. Note that in most of the 
phase space the displacement of the transition temperature due to a nonvanishing $\eps''$ is 
much smaller than that due to $\eps'$; in spatially nonuniform fields far from the critical 
composition, the demixing transition is well described by a linear constitutive relation 
$\eps(\phi)$.

\subsection{Demixing transitions for $\phi>\phi_c$}

\begin{figure}[!bt]
%fig : /users/sela/work/efips/newQ/efips_#2_figs
%plot : plot_efips_prof_high.m
\centerline{\includegraphics[width=3.5in,clip]{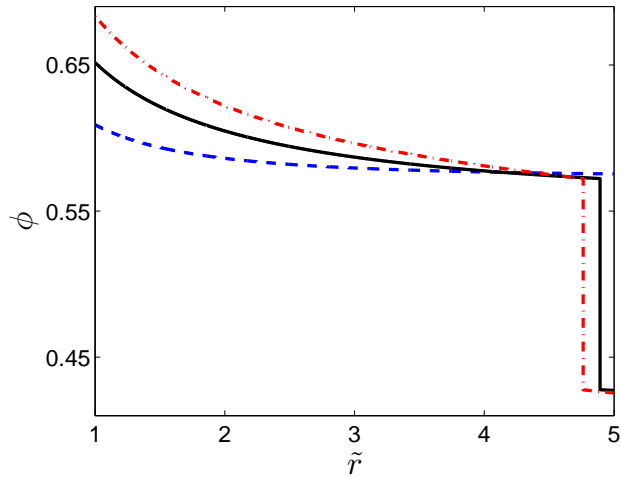}}
\caption{Equilibrium profiles $\phi(\tr)$ for a closed wedge system with an average 
composition $\phi_0=0.58$ larger than $\phi_c$. Dashed line: $M_{{\rm w}}=1\times10^{-3}$ 
smaller than $M_{{\rm w}}^*$, and $\phi(\tr)$ is smoothly varying. Solid line: $M_{{\rm 
w}}=4\times10^{-3}$ is large enough to induce phase separation. Dash-dot line: $M_{{\rm 
w}}=8\times10^{-3}$, the separation interface moves to a smaller radius.  We used 
$T=0.993T_c$.}
\label{fig_high_type}
\end{figure}

Since the electric field breaks the symmetry of the free energy with respect to composition 
($\phi \rightarrow 1-\phi $), the full stability diagram is asymmetric with respect to 
$\phi-\phi_c$. \figref{fig_sd2} (a) shows that in an open system, phase separation does not 
occur when $\phi_0>\phi_c$. In \figref{fig_sd2} (b) there are unstable compositions $\phi_0$ 
such that $\phi_0>\phi_c$. This is an important feature of the stability diagram in closed 
systems. Here, the dielectrophoretic force that pulls the high dielectric constant liquid 
toward the electrode creates a depletion in the region where electric field is low, and phase 
separation will occur if the composition at this region is close to the binodal composition. 
The stability curves in \figref{fig_sd2} (b) show that higher values of potential or charge 
are required for demixing when $\phi_0>\phi_c$. Indeed, the ratio of electrostatic energies at 
the inner and outer radii is $\simeq \tR_2^2/\tR_1^2$.

\begin{figure}[!tb]
\begin{center}.
%fig : /users/sela/work/efips/newQ/efips_#2_figs
%plot : plot_efips_rt_high.m
\includegraphics[width=3in,clip]{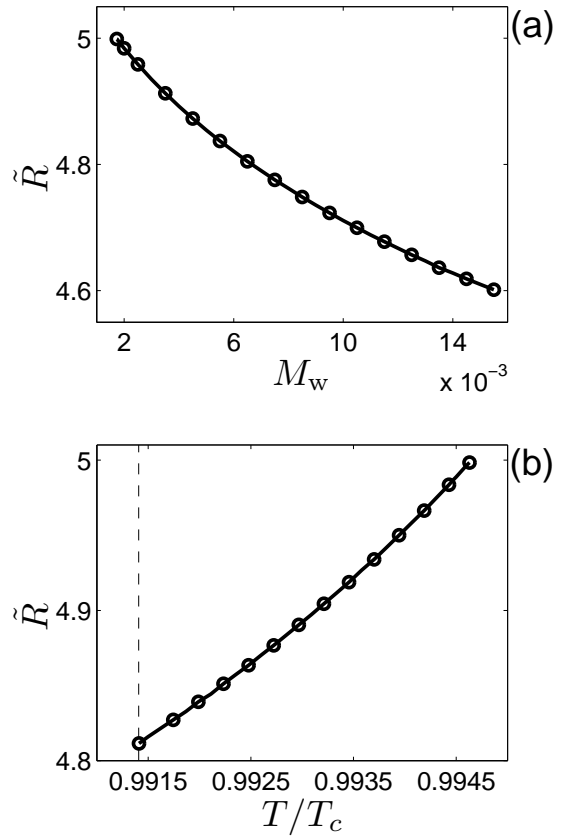}
\end{center}
\caption{(a) Interface location $\tR$ for a closed wedge system with an average composition 
$\phi_0=0.58>\phi_c$ as a function of $M_{{\rm w}}$ at $T=0.993T_c$. The minimal value of 
$\tR$ is 
$\tR_{\infty}=3.86$ given by \eqref{eq:rinf}. (b) $\tR$ as a function of $T$ at 
$M_{{\rm w}}=4\times10^{-3}$. Dashed line is $T_t$.}
\label{fig_high_Rt}
\end{figure}

Examples of this behavior are shown in \figref{fig_high_type}. When $\phi>\phi_c$, and $M$ is 
small (dashed line) $\phi(\tr)$ varies smoothly, exhibiting the usual dielectrophoretic 
behavior. However, if $M$ is sufficiently large, that is $M>M^{*}$, the long range 
dielectrophoretic force gives rise to a phase-separation transition near $\tR_2$ (solid line). 
If $M$ is further increased the phase-separation interface moves to smaller radii (dash-dot 
line).

\figref{fig_high_Rt} shows how $\tR$ varies with $M$ and $T$ in the wedge geometry for 
$\phi>\phi_c$. At $M^{*}$, the interface appears where the field is minimal, i.e at $\tR_2=5$, 
see \figref{fig_high_Rt} (a). An increase in $M$ decreases $\tR$ and the volume of the 
high-$\phi$ domain. \figref{fig_high_Rt} (b) shows that at fixed $M$, $\tR$ decreases with $T$ 
and attains its minimal value at $T_t$.  The minimal value of $\tR$ is $\tR_{\infty}$ given by 
\eqref{eq:rinf}.

When $\eps''\neq0$, the stability curves in \figref{fig_sd2} (b) have another kink at 
$\phi>\phi_c$ at a temperature we denote $T_{k,2}$. From the previous discussion, $T_{k,2}$ is 
obtained by replacing $\tR_1$ by $\tR_2$ in \eqref{eqn_pTk}. In the wedge geometry we thus find
\begin{eqnarray}\label{eqn_tk2}
\frac{T_{k,2}}{T_c}=1+\frac{M_{{\rm w}}\eps''}{\tR_2^2}.
\end{eqnarray} 
Hence, $T_c-T_{k,2}$ is smaller than $T_c-T_{k,1}$ by a factor of $\tR_2^2$.

\section{Conclusions}

We present a systematic study of electric field induced phase-separation transitions in binary
mixtures on the mean field level. The behavior described by us complements the findings
of Onuki and co-workers \cite{onuki:021506,onuki:3143,onuki_jpcb2009} and
\textcite{andelman_jpcb2009}
who mainly focused on effects above the
critical temperature or on miscible liquids.
The transitions should occur in any bistable system with a
dielectric mismatch sufficiently close to the transition temperature, e.g a vapor phase close
to coexistence with its liquid.
The differences between closed and open systems and between the three geometries are
highlighted. The stability diagrams in the temperature-composition plane are given. These
diagrams show that the change in the transition temperature is much larger in nonuniform fields
than in uniform fields. 
% The phase-separation transition is captured by a linear constitutive
% relation. 
We describe the special temperatures $T_{k,1}$ and $T_{k,2}$ where the stability
diagram has a ``kink''. It should be emphasized that the location of the phase separation
interface is not restricted to the vicinity of the confining surfaces, and can appear at a
finite location $\tR$ in the range of temperatures $T_{k,1}<T<T_{k,2}$ as described above in
\figref{fig_sd2}. In other geometries, too, the demixing interface can be created far from any
surface, e.g in quadrupolar electrode array.

In order to test our predictions we suggest the following experiments. Consider 
a wedge
capacitor partially immersed in a binary liquid mixture near the coexistence
temperature. Neutron reflectometry may be used to probe
directly the composition profile \cite{Bowers2007} and to determine the critical voltage for
demixing. Here we expect $\tR$ to be proportional to the voltage difference across the
electrodes. A different experiment can be realized by suspending a conducting
wire in a dilute vapor phase of a pure substance near the coexistence temperature
with the liquid. 
When a potential $V$ is applied to the wire, its fundamental
frequency of vibration perpendicular to its length changes from $f_0$ to $f_{V}$ because
a liquid layer condenses around the wire. We expect that the ratio $(f_0/f_{V})^2$ should
be a linear function of $V^2$. Alternatively, one can measure directly the change in the wire
mass $\Delta m$ using a microbalance -- $\Delta m$ should depend linearly on $V^2$. 
In these experiments, care must be taken to subtract wetting or confinement effects due to the
electrodes.

Capillary condensation has been described
\cite{Dobbs1992,PhysRevE.62.5324,PhysRevLett.54.2123,Narayanan1993,Koehler1997} 
for colloidal suspensions in binary mixtures. According to our work, charged colloidal
suspensions can flocculate or be stabilized due to the formation of liquid layers around the
colloids. The liquid layer may also influence the interaction between a colloid and a flat
surface both far and close to the critical point \cite{Bechinger2008}.

Our results may also be measurable in Surface Force Balance and Atomic Force Microscope
experiments. In such apparatus, a bridging transition has been observed in binary mixtures
\cite{Wennerstrom1998} and explained by capillary
forces \cite{andrienko2004,Olsson2004}. Since in most cases the surfaces are charged, a
capillary bridge could be the result of the merging of two field-induced layers, leading to
long range attractive forces between the surfaces.

In the current work we neglect a
$(\nabla \phi)^2$ in the mixing free energy $f_m$, since the system size was relatively
large and the electrostatic energy acts throughout the whole system volume, and therefore
is dominant \cite{onuki_book}. We have verified that inclusion of such term leads to
smoothing of composition profiles but otherwise to no other noticeable changes
in the figures presented. In order to isolate the electric field effect, we have also not
considered any direct short- or long-range interactions with the confining surfaces. In
a future study it would be very interesting to relax these assumptions and to look at ever
smaller systems. Here we are intrigued by the possibility to find qualitative, and not
just quantitative, differences from our current profiles and diagrams.

\section*{ACKNOWLEDGMENTS}

We acknowledge useful discussions with L. Leibler, D. Andelman, H. Diamant, and J. Klein, and
critical comments from K. Binder.
This research was supported by the Israel Science
foundation (ISF) grant no. 284/05, by the German Israeli Foundation (GIF) grant
no. 2144-1636.10/2006, and by the COST European program P21 ``The Physics of Drops''.

\bibliography{refs}

\begin{thebibliography}{38}
\expandafter\ifx\csname natexlab\endcsname\relax\def\natexlab#1{#1}\fi
\expandafter\ifx\csname bibnamefont\endcsname\relax
  \def\bibnamefont#1{#1}\fi
\expandafter\ifx\csname bibfnamefont\endcsname\relax
  \def\bibfnamefont#1{#1}\fi
\expandafter\ifx\csname citenamefont\endcsname\relax
  \def\citenamefont#1{#1}\fi
\expandafter\ifx\csname url\endcsname\relax
  \def\url#1{\texttt{#1}}\fi
\expandafter\ifx\csname urlprefix\endcsname\relax\def\urlprefix{URL }\fi
\providecommand{\bibinfo}[2]{#2}
\providecommand{\eprint}[2][]{\url{#2}}

\bibitem[{\citenamefont{Landau and Lifshitz}(1957)}]{landau}
\bibinfo{author}{\bibfnamefont{L.~D.} \bibnamefont{Landau}} \bibnamefont{and}
  \bibinfo{author}{\bibfnamefont{E.~M.} \bibnamefont{Lifshitz}},
  \emph{\bibinfo{title}{Elektrodinamika Sploshnykh Sred Chap. II, Sec. 18,
  Problem 1}} (\bibinfo{publisher}{Nauka}, \bibinfo{address}{Moscow},
  \bibinfo{year}{1957}).

\bibitem[{\citenamefont{Onuki}(1995)}]{onuki1}
\bibinfo{author}{\bibfnamefont{A.}~\bibnamefont{Onuki}},
  \bibinfo{journal}{Europhys. Lett.} \textbf{\bibinfo{volume}{29}},
  \bibinfo{pages}{611} (\bibinfo{year}{1995}).

\bibitem[{\citenamefont{Debye and Kleboth}(1965)}]{debye}
\bibinfo{author}{\bibfnamefont{P.}~\bibnamefont{Debye}} \bibnamefont{and}
  \bibinfo{author}{\bibfnamefont{K.}~\bibnamefont{Kleboth}},
  \bibinfo{journal}{J. Chem. Phys.} \textbf{\bibinfo{volume}{42}},
  \bibinfo{pages}{3155} (\bibinfo{year}{1965}).

\bibitem[{\citenamefont{Beaglehole}(1981)}]{beaglehole}
\bibinfo{author}{\bibfnamefont{D.}~\bibnamefont{Beaglehole}},
  \bibinfo{journal}{J. Chem. Phys.} \textbf{\bibinfo{volume}{74}},
  \bibinfo{pages}{5251} (\bibinfo{year}{1981}).

\bibitem[{\citenamefont{Orzechowski}(1999)}]{orzechowski}
\bibinfo{author}{\bibfnamefont{K.}~\bibnamefont{Orzechowski}},
  \bibinfo{journal}{Chem. Phys.} \textbf{\bibinfo{volume}{240}},
  \bibinfo{pages}{275} (\bibinfo{year}{1999}).

\bibitem[{\citenamefont{Wirtz and Fuller}(1993)}]{wirtz}
\bibinfo{author}{\bibfnamefont{D.}~\bibnamefont{Wirtz}} \bibnamefont{and}
  \bibinfo{author}{\bibfnamefont{G.~G.} \bibnamefont{Fuller}},
  \bibinfo{journal}{Phys. Rev. Lett.} \textbf{\bibinfo{volume}{71}},
  \bibinfo{pages}{2236} (\bibinfo{year}{1993}).

\bibitem[{\citenamefont{Reich and Gordon}(1979)}]{reich}
\bibinfo{author}{\bibfnamefont{S.}~\bibnamefont{Reich}} \bibnamefont{and}
  \bibinfo{author}{\bibfnamefont{J.~M.} \bibnamefont{Gordon}},
  \bibinfo{journal}{J. Pol. Sci.: Pol. Phys.} \textbf{\bibinfo{volume}{17}},
  \bibinfo{pages}{371} (\bibinfo{year}{1979}).

\bibitem[{\citenamefont{G{\'a}bor and Szalai}(2008)}]{Szalai2008}
\bibinfo{author}{\bibfnamefont{A.}~\bibnamefont{G{\'a}bor}} \bibnamefont{and}
  \bibinfo{author}{\bibfnamefont{I.}~\bibnamefont{Szalai}},
  \bibinfo{journal}{Molecular Physics} \textbf{\bibinfo{volume}{106}},
  \bibinfo{pages}{801} (\bibinfo{year}{2008}), ISSN \bibinfo{issn}{0026-8976}.

\bibitem[{\citenamefont{Onuki}(2006)}]{onuki:021506}
\bibinfo{author}{\bibfnamefont{A.}~\bibnamefont{Onuki}},
  \bibinfo{journal}{Phys. Rev. E} \textbf{\bibinfo{volume}{73}},
  \bibinfo{eid}{021506} (\bibinfo{year}{2006}).

\bibitem[{\citenamefont{Schoberth et~al.}(2009)\citenamefont{Schoberth,
  Schmidt, Schindler, and Boker}}]{boker_mm2009}
\bibinfo{author}{\bibfnamefont{H.~G.} \bibnamefont{Schoberth}},
  \bibinfo{author}{\bibfnamefont{K.}~\bibnamefont{Schmidt}},
  \bibinfo{author}{\bibfnamefont{K.~A.} \bibnamefont{Schindler}},
  \bibnamefont{and} \bibinfo{author}{\bibfnamefont{A.}~\bibnamefont{Boker}},
  \bibinfo{journal}{Macromolecules} \textbf{\bibinfo{volume}{42}},
  \bibinfo{pages}{3433} (\bibinfo{year}{2009}).

\bibitem[{\citenamefont{Tsori et~al.}(2006)\citenamefont{Tsori, Andelman, Lin,
  and Schick}}]{tsori_mm2006}
\bibinfo{author}{\bibfnamefont{Y.}~\bibnamefont{Tsori}},
  \bibinfo{author}{\bibfnamefont{D.}~\bibnamefont{Andelman}},
  \bibinfo{author}{\bibfnamefont{C.-Y.} \bibnamefont{Lin}}, \bibnamefont{and}
  \bibinfo{author}{\bibfnamefont{M.}~\bibnamefont{Schick}},
  \bibinfo{journal}{Macromolecules} \textbf{\bibinfo{volume}{39}},
  \bibinfo{pages}{289} (\bibinfo{year}{2006}).

\bibitem[{\citenamefont{Tsori}(2009)}]{tsori_rmp2009}
\bibinfo{author}{\bibfnamefont{Y.}~\bibnamefont{Tsori}}, \bibinfo{journal}{Rev.
  Mod. Phys.} \textbf{\bibinfo{volume}{81}}, \bibinfo{pages}{1471}
  (\bibinfo{year}{2009}).

\bibitem[{com({\natexlab{a}})}]{comment1}
\emph{\bibinfo{title}{{\rm It has been argued by Stepanow and co-workers that
  fluctuation effects near the critical point change this behavior
  \cite{stepanow_mm2007,stepanow_pre2009}.}}}

\bibitem[{\citenamefont{Gunkel et~al.}(2007)\citenamefont{Gunkel, Stepanow,
  Thurn-Albrecht, and Trimper}}]{stepanow_mm2007}
\bibinfo{author}{\bibfnamefont{I.}~\bibnamefont{Gunkel}},
  \bibinfo{author}{\bibfnamefont{S.}~\bibnamefont{Stepanow}},
  \bibinfo{author}{\bibfnamefont{T.}~\bibnamefont{Thurn-Albrecht}},
  \bibnamefont{and} \bibinfo{author}{\bibfnamefont{S.}~\bibnamefont{Trimper}},
  \bibinfo{journal}{Macromolecules} \textbf{\bibinfo{volume}{40}},
  \bibinfo{pages}{2186} (\bibinfo{year}{2007}).

\bibitem[{\citenamefont{Stepanow and Thurn-Albrecht}(2009)}]{stepanow_pre2009}
\bibinfo{author}{\bibfnamefont{S.}~\bibnamefont{Stepanow}} \bibnamefont{and}
  \bibinfo{author}{\bibfnamefont{T.}~\bibnamefont{Thurn-Albrecht}},
  \bibinfo{journal}{Phys. Rev. E} \textbf{\bibinfo{volume}{79}},
  \bibinfo{pages}{041104} (\bibinfo{year}{2009}).

\bibitem[{\citenamefont{Tsori et~al.}(2004)\citenamefont{Tsori, Tournilhac, and
  Leibler}}]{efdemix}
\bibinfo{author}{\bibfnamefont{Y.}~\bibnamefont{Tsori}},
  \bibinfo{author}{\bibfnamefont{F.}~\bibnamefont{Tournilhac}},
  \bibnamefont{and} \bibinfo{author}{\bibfnamefont{L.}~\bibnamefont{Leibler}},
  \bibinfo{journal}{Nature} \textbf{\bibinfo{volume}{430}},
  \bibinfo{pages}{544} (\bibinfo{year}{2004}).

\bibitem[{\citenamefont{Marcus et~al.}(2008)\citenamefont{Marcus, Samin, and
  Tsori}}]{efips_jcp1}
\bibinfo{author}{\bibfnamefont{G.}~\bibnamefont{Marcus}},
  \bibinfo{author}{\bibfnamefont{S.}~\bibnamefont{Samin}}, \bibnamefont{and}
  \bibinfo{author}{\bibfnamefont{Y.}~\bibnamefont{Tsori}}, \bibinfo{journal}{J.
  Chem. Phys.} \textbf{\bibinfo{volume}{129}}, \bibinfo{eid}{061101}
  (\bibinfo{year}{2008}).

\bibitem[{\citenamefont{Landau et~al.}(1984)\citenamefont{Landau, Lifshitz, and
  Pitaevskii}}]{landau2}
\bibinfo{author}{\bibfnamefont{L.~D.} \bibnamefont{Landau}},
  \bibinfo{author}{\bibfnamefont{E.~M.} \bibnamefont{Lifshitz}},
  \bibnamefont{and} \bibinfo{author}{\bibfnamefont{L.~P.}
  \bibnamefont{Pitaevskii}}, \emph{\bibinfo{title}{Electrodynamics of
  Continuous Media}} (\bibinfo{publisher}{Butterworth-Heinemann},
  \bibinfo{address}{Amsterdam}, \bibinfo{year}{1984}), \bibinfo{edition}{2nd}
  ed.

\bibitem[{\citenamefont{Onuki}(2004{\natexlab{a}})}]{onuki_nato2004}
\bibinfo{author}{\bibfnamefont{A.}~\bibnamefont{Onuki}},
  \emph{\bibinfo{title}{Nonlinear dielectric phenomena in complex liquids}}
  (\bibinfo{publisher}{Kluwer Academic, Dordrecht},
  \bibinfo{year}{2004}{\natexlab{a}}).

\bibitem[{\citenamefont{Onuki and Kitamura}(2004)}]{onuki:3143}
\bibinfo{author}{\bibfnamefont{A.}~\bibnamefont{Onuki}} \bibnamefont{and}
  \bibinfo{author}{\bibfnamefont{H.}~\bibnamefont{Kitamura}},
  \bibinfo{journal}{J. Chem. Phys.} \textbf{\bibinfo{volume}{121}},
  \bibinfo{pages}{3143} (\bibinfo{year}{2004}).

\bibitem[{\citenamefont{Onuki}(2004{\natexlab{b}})}]{onuki_book}
\bibinfo{author}{\bibfnamefont{A.}~\bibnamefont{Onuki}},
  \emph{\bibinfo{title}{Phase transition dynamics}}
  (\bibinfo{publisher}{Cambridge University Press},
  \bibinfo{year}{2004}{\natexlab{b}}).

\bibitem[{\citenamefont{Doi}(1996)}]{doi}
\bibinfo{author}{\bibfnamefont{M.}~\bibnamefont{Doi}},
  \emph{\bibinfo{title}{Introduction to polymer physics}}
  (\bibinfo{publisher}{Oxford University Press}, \bibinfo{address}{Oxford, UK},
  \bibinfo{year}{1996}).

\bibitem[{\citenamefont{Akhadov}(1981)}]{Akhadov}
\bibinfo{author}{\bibfnamefont{Y.~Y.} \bibnamefont{Akhadov}},
  \emph{\bibinfo{title}{Dielectric properties of binary solutions}}
  (\bibinfo{publisher}{Oxford : Pergamon Press}, \bibinfo{year}{1981}).

\bibitem[{\citenamefont{Sen et~al.}(1992)\citenamefont{Sen, Anicich, and
  Arakelian}}]{Sen1992}
\bibinfo{author}{\bibfnamefont{A.~D.} \bibnamefont{Sen}},
  \bibinfo{author}{\bibfnamefont{V.~G.} \bibnamefont{Anicich}},
  \bibnamefont{and}
  \bibinfo{author}{\bibfnamefont{T.}~\bibnamefont{Arakelian}},
  \bibinfo{journal}{J. Phys. D: Appl. Phys.} \textbf{\bibinfo{volume}{25}},
  \bibinfo{pages}{616} (\bibinfo{year}{1992}).

\bibitem[{com({\natexlab{b}})}]{comment2}
\emph{\bibinfo{title}{{\rm Strictly speaking, the spinodal compositions
  $\phi_s$ are ``forbidden'', because the spinodal is the locus of critical
  points, and thus critical fluctuations are expected to be important. The
  current mean-field treaties therefore is not expected to hold when
  $\phi\simeq\phi_s$.}}}

\bibitem[{\citenamefont{Landau and Lifshitz}(1980)}]{landau3}
\bibinfo{author}{\bibfnamefont{L.~D.} \bibnamefont{Landau}} \bibnamefont{and}
  \bibinfo{author}{\bibfnamefont{E.~M.} \bibnamefont{Lifshitz}},
  \emph{\bibinfo{title}{Statistical Physics Part 1 \S 144}}
  (\bibinfo{publisher}{Butterworth-Heinemann}, \bibinfo{address}{Amsterdam},
  \bibinfo{year}{1980}), \bibinfo{edition}{3rd} ed.

\bibitem[{\citenamefont{Onuki and Okamoto}(2009)}]{onuki_jpcb2009}
\bibinfo{author}{\bibfnamefont{A.}~\bibnamefont{Onuki}} \bibnamefont{and}
  \bibinfo{author}{\bibfnamefont{R.}~\bibnamefont{Okamoto}},
  \bibinfo{journal}{J. Phys. Chem. B} \textbf{\bibinfo{volume}{113}},
  \bibinfo{pages}{3988} (\bibinfo{year}{2009}).

\bibitem[{\citenamefont{Ben-Yaakov et~al.}(2009)\citenamefont{Ben-Yaakov,
  Andelman, Harries, and Podgornik}}]{andelman_jpcb2009}
\bibinfo{author}{\bibfnamefont{D.}~\bibnamefont{Ben-Yaakov}},
  \bibinfo{author}{\bibfnamefont{D.}~\bibnamefont{Andelman}},
  \bibinfo{author}{\bibfnamefont{D.}~\bibnamefont{Harries}}, \bibnamefont{and}
  \bibinfo{author}{\bibfnamefont{R.}~\bibnamefont{Podgornik}},
  \bibinfo{journal}{J. Phys. Chem. B} \textbf{\bibinfo{volume}{113}},
  \bibinfo{pages}{6001} (\bibinfo{year}{2009}).

\bibitem[{\citenamefont{Bowers et~al.}(2007)\citenamefont{Bowers, Zarbakhsh,
  McLure, Webster, Steitz, and Christenson}}]{Bowers2007}
\bibinfo{author}{\bibfnamefont{J.}~\bibnamefont{Bowers}},
  \bibinfo{author}{\bibfnamefont{A.}~\bibnamefont{Zarbakhsh}},
  \bibinfo{author}{\bibfnamefont{I.~A.} \bibnamefont{McLure}},
  \bibinfo{author}{\bibfnamefont{J.~R.~P.} \bibnamefont{Webster}},
  \bibinfo{author}{\bibfnamefont{R.}~\bibnamefont{Steitz}}, \bibnamefont{and}
  \bibinfo{author}{\bibfnamefont{H.~K.} \bibnamefont{Christenson}},
  \bibinfo{journal}{J. Phys. Chem. C} \textbf{\bibinfo{volume}{111}},
  \bibinfo{pages}{5568} (\bibinfo{year}{2007}).

\bibitem[{\citenamefont{Dobbs and Yeomans}(1992)}]{Dobbs1992}
\bibinfo{author}{\bibfnamefont{H.~T.} \bibnamefont{Dobbs}} \bibnamefont{and}
  \bibinfo{author}{\bibfnamefont{J.~M.} \bibnamefont{Yeomans}},
  \bibinfo{journal}{J. Phys.: Condens. Matter} \textbf{\bibinfo{volume}{4}},
  \bibinfo{pages}{10133} (\bibinfo{year}{1992}).

\bibitem[{\citenamefont{Bauer et~al.}(2000)\citenamefont{Bauer, Bieker, and
  Dietrich}}]{PhysRevE.62.5324}
\bibinfo{author}{\bibfnamefont{C.}~\bibnamefont{Bauer}},
  \bibinfo{author}{\bibfnamefont{T.}~\bibnamefont{Bieker}}, \bibnamefont{and}
  \bibinfo{author}{\bibfnamefont{S.}~\bibnamefont{Dietrich}},
  \bibinfo{journal}{Phys. Rev. E} \textbf{\bibinfo{volume}{62}},
  \bibinfo{pages}{5324} (\bibinfo{year}{2000}).

\bibitem[{\citenamefont{Beysens and Est{\`e}ve}(1985)}]{PhysRevLett.54.2123}
\bibinfo{author}{\bibfnamefont{D.}~\bibnamefont{Beysens}} \bibnamefont{and}
  \bibinfo{author}{\bibfnamefont{D.}~\bibnamefont{Est{\`e}ve}},
  \bibinfo{journal}{Phys. Rev. Lett.} \textbf{\bibinfo{volume}{54}},
  \bibinfo{pages}{2123} (\bibinfo{year}{1985}).

\bibitem[{\citenamefont{Narayanan et~al.}(1993)\citenamefont{Narayanan, Kumar,
  Gopal, Beysens, Guenoun, and Zalczer}}]{Narayanan1993}
\bibinfo{author}{\bibfnamefont{T.}~\bibnamefont{Narayanan}},
  \bibinfo{author}{\bibfnamefont{A.}~\bibnamefont{Kumar}},
  \bibinfo{author}{\bibfnamefont{E.~S.~R.} \bibnamefont{Gopal}},
  \bibinfo{author}{\bibfnamefont{D.}~\bibnamefont{Beysens}},
  \bibinfo{author}{\bibfnamefont{P.}~\bibnamefont{Guenoun}}, \bibnamefont{and}
  \bibinfo{author}{\bibfnamefont{G.}~\bibnamefont{Zalczer}},
  \bibinfo{journal}{Phys. Rev. E} \textbf{\bibinfo{volume}{48}},
  \bibinfo{pages}{1989} (\bibinfo{year}{1993}).

\bibitem[{\citenamefont{Koehler and Kaler}(1997)}]{Koehler1997}
\bibinfo{author}{\bibfnamefont{R.~D.} \bibnamefont{Koehler}} \bibnamefont{and}
  \bibinfo{author}{\bibfnamefont{E.~W.} \bibnamefont{Kaler}},
  \bibinfo{journal}{Langmuir} \textbf{\bibinfo{volume}{13}},
  \bibinfo{pages}{2463} (\bibinfo{year}{1997}).

\bibitem[{\citenamefont{Hertlein et~al.}(2008)\citenamefont{Hertlein, Helden,
  Gambassi, Dietrich, and Bechinger}}]{Bechinger2008}
\bibinfo{author}{\bibfnamefont{C.}~\bibnamefont{Hertlein}},
  \bibinfo{author}{\bibfnamefont{L.}~\bibnamefont{Helden}},
  \bibinfo{author}{\bibfnamefont{A.}~\bibnamefont{Gambassi}},
  \bibinfo{author}{\bibfnamefont{S.}~\bibnamefont{Dietrich}}, \bibnamefont{and}
  \bibinfo{author}{\bibfnamefont{C.}~\bibnamefont{Bechinger}},
  \bibinfo{journal}{Nature} \textbf{\bibinfo{volume}{451}},
  \bibinfo{pages}{172} (\bibinfo{year}{2008}).

\bibitem[{\citenamefont{Wennerstrom et~al.}(1998)\citenamefont{Wennerstrom,
  Thuresson, Linse, and Freyssingeas}}]{Wennerstrom1998}
\bibinfo{author}{\bibfnamefont{H.}~\bibnamefont{Wennerstrom}},
  \bibinfo{author}{\bibfnamefont{K.}~\bibnamefont{Thuresson}},
  \bibinfo{author}{\bibfnamefont{P.}~\bibnamefont{Linse}}, \bibnamefont{and}
  \bibinfo{author}{\bibfnamefont{E.}~\bibnamefont{Freyssingeas}},
  \bibinfo{journal}{Langmuir} \textbf{\bibinfo{volume}{14}},
  \bibinfo{pages}{5664} (\bibinfo{year}{1998}).

\bibitem[{\citenamefont{Andrienko et~al.}(2004)\citenamefont{Andrienko,
  Patricio, and Vinogradova}}]{andrienko2004}
\bibinfo{author}{\bibfnamefont{D.}~\bibnamefont{Andrienko}},
  \bibinfo{author}{\bibfnamefont{P.}~\bibnamefont{Patricio}}, \bibnamefont{and}
  \bibinfo{author}{\bibfnamefont{O.~I.} \bibnamefont{Vinogradova}},
  \bibinfo{journal}{J. Chem. Phys.} \textbf{\bibinfo{volume}{121}},
  \bibinfo{pages}{4414} (\bibinfo{year}{2004}).

\bibitem[{\citenamefont{Olsson et~al.}(2004)\citenamefont{Olsson, Linse, and
  Piculell}}]{Olsson2004}
\bibinfo{author}{\bibfnamefont{M.}~\bibnamefont{Olsson}},
  \bibinfo{author}{\bibfnamefont{P.}~\bibnamefont{Linse}}, \bibnamefont{and}
  \bibinfo{author}{\bibfnamefont{L.}~\bibnamefont{Piculell}},
  \bibinfo{journal}{Langmuir} \textbf{\bibinfo{volume}{20}},
  \bibinfo{pages}{1611} (\bibinfo{year}{2004}).

\end{thebibliography}

\end{document}